\RequirePackage{fixltx2e} 
\documentclass[aip,jcp,preprint,floatfix]{revtex4-1}
\usepackage{graphicx} 
\usepackage{color}
\usepackage{dcolumn}
\usepackage{amsmath,amsfonts} %
\usepackage{braket} %

\newcommand{\xx}{\mathbf{x}} %
\newcommand{\qq}{\mathbf{q}} %
\newcommand{\pp}{\mathbf{p}} %
\newcommand{\eq}[1]{Eq.~(\ref{#1})} %
\newcommand{\eqs}[1]{Eqs.~(\ref{#1})} %
\newcommand{\fig}[1]{Fig.~\ref{#1}} %
\newcommand{\bea}{\begin{eqnarray}}
\newcommand{\eea}{\end{eqnarray}}
                            
\begin{document}

\title{Quantum Nonadiabatic Cloning of Entangled Coherent States}
\author{Artur F. Izmaylov} %
\email{artur.izmaylov@utoronto.ca}
\affiliation{Department of Physical and Environmental Sciences, University of Toronto Scarborough, Toronto, Ontario, M1C 1A4, Canada} %
\affiliation{Chemical Physics Theory Group, Department of Chemistry, University of Toronto, Toronto, Ontario M5S 3H6, Canada} %
\author{Lo{\"i}c Joubert-Doriol} %
\affiliation{Department of Physical and Environmental Sciences, University of Toronto Scarborough, Toronto, Ontario, M1C 1A4, Canada} %
\affiliation{Chemical Physics Theory Group, Department of Chemistry, University of Toronto, Toronto, Ontario M5S 3H6, Canada} %


\begin{abstract}
We propose a systematic approach to the basis set extension for nonadiabatic dynamics of 
entangled combination of nuclear coherent states (CSs) evolving 
according to the time-dependent variational principle (TDVP). 
TDVP provides a rigorous framework for fully quantum nonadiabatic dynamics of closed systems,
however, quality of results strongly depends on available basis functions. Starting with 
a single nuclear CS replicated vertically on all electronic states, our approach clones this function 
when replicas of the CS on different electronic states experience increasingly different forces. 
Created clones move away from each other (decohere) extending the basis set. 
To determine a moment for cloning we introduce generalized forces based  
on derivatives that maximally contribute to a variation of the total quantum action  
and thus account for entanglement of all basis functions. 
\end{abstract}


\maketitle



The time-dependent variational principle (TDVP)\cite{Book/Kramer:1981,
Dirac:1958/Book,Book/Frenkel:1934} provides variationally optimal 
equations of motion (EOM) for the system wave-function specified by a certain ansatz. 
The TDVP allows one to model the quantum nuclear wave-function in a 
computationally efficient way for both adiabatic and nonadiabatic nuclear dynamics in molecules. 
There are two main popular forms of the nuclear wave-function: 
1) originating from the multi-configuration time-dependent Hartree (MCTDH) method\cite{mey90:73,Wang:2003/jcp/1289,mctdh} and its multilayer generalizations,\cite{Wang:2003fu,Manthe:2008ev} 2) based on frozen-width gaussians, 
\cite{Yang:2009ja,BenNun:2002tx,Shalashilin:2009/JCP/244101,Burghardt:2008iz,Worth:2008/MP/2077,Worth:2004/FD/307,Izmaylov:2013fe,Makhov:2014/jcp/054110,Fernandez:2016/pccp/10028}
which are moving either classically\cite{Yang:2009ja,BenNun:2002tx,Shalashilin:2009/JCP/244101,Makhov:2014/jcp/054110,Fernandez:2016/pccp/10028} or quantum-mechanically\cite{Burghardt:2008iz,Worth:2008/MP/2077,Worth:2004/FD/307,Izmaylov:2013fe}. The latter ansatz, due to locality of involved basis functions, is very well suited 
 to be used in conjunction with the on-the-fly solution of the electronic structure problem.
\cite{BenNun:2002tx,Shalashilin:2009/JCP/244101,Fernandez:2016/pccp/10028}

\begin{figure}
  \centering
  \includegraphics[width=0.48\textwidth]{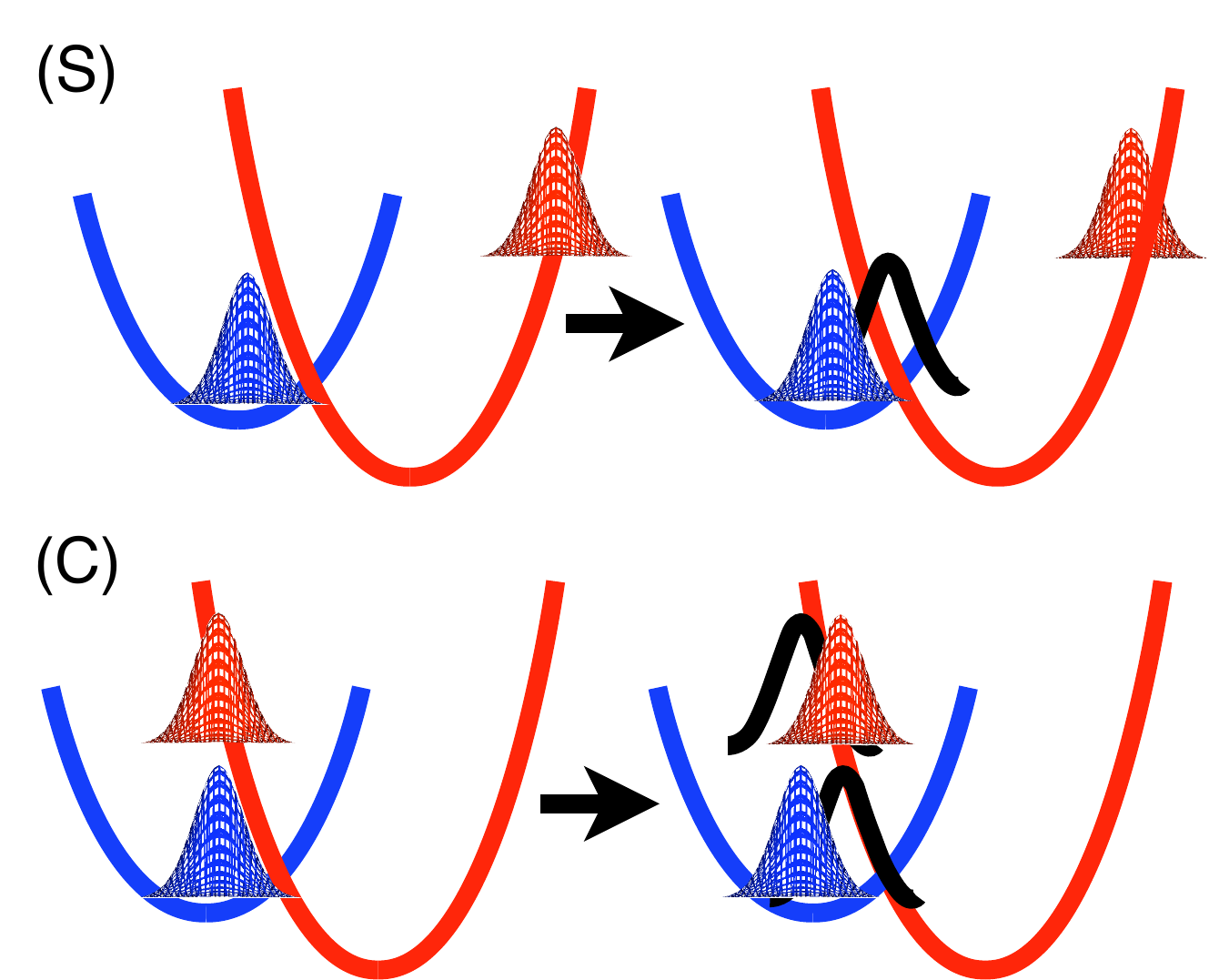}
  \caption{Illustration of the spawning (S) and cloning (C) procedures.}
  \label{fig:spcl}
\end{figure}

The main practical difficulty for any dynamical method based on the TDVP 
is basis set limitation. If we consider nonadiabatic dynamics using a linear combination of 
frozen-width gaussians 
\bea\label{eq:wfms}
\ket{\Psi(t)} = \sum_{I=1}^{N_G}\sum_{s=1}^{N_s} C_{I}^{(s)}(t) \ket{G_{I}^{(s)}(t)} \ket{\phi_{s}},
\eea
where  $C_{I}^{(s)}$ are time-dependent coefficients (amplitudes), 
indices $s$ and $I$ enumerate
electronic states $\ket{\phi_{s}}$ and gaussians $\ket{G_{I}^{(s)}}$, respectively, 
the population transfer between electronic states can only take place when
gaussians located on different electronic states 
have significant overlap in nuclear degrees of freedom (DOF), 
$\Braket{G_I^{(s)}\Big{\vert} G_J^{(s')}}\gg 0$.
However, considering localized nature of gaussians and that different electronic surfaces provide 
 different forces in the same area of nuclear 
geometry, these overlaps generally quickly decay along the dynamics. This decoherence process 
artificially reduces the electronic population transfer. 
To address this issue,  
the spawning technique was introduced for a linear combination of frozen-width gaussians whose 
parameters evolved classically while the amplitudes were propagated quantum-mechanically.\cite{Yang:2009ja,BenNun:2002tx}
If a gaussian arrives at a region of strong coupling between electronic states and there is no 
gaussian on the other state to interact with it, the spawning algorithm creates the counterpart 
needed for population exchange (\fig{fig:spcl}S).
This consideration may seem {\it ad hoc} and does not account for the fact that
each gaussian basis function is a part of the total nuclear wave-function. 
However, the spawning approach can be also rigorously 
introduced using time-dependent perturbation theory\cite{Izmaylov:2013fe} that takes the total 
wave-function into account and provides a route for dynamical basis set extension. 
This perturbative spawning has been extended to the fully quantum propagation 
schemes such as the variational multiconfiguration gaussian (vMCG) method 
where gaussian dynamics has highly entangled quantum character.\cite{Izmaylov:2013fe} 

Alternatively, one can approach the problem of population transfer in TDVP based nonadiabatic 
dynamics by introducing a nuclear basis with the condition 
\bea\label{eq:constr}
\ket{G_{I}^{(s)}} = \ket{G_{I}^{(s')}} = \ket{G_{I}},~s\ne s'.
\eea
This condition will ensure the maximum overlap between gaussians on 
different electronic states $\Braket{G_I^{(s)}\Big{\vert} G_I^{(s')}}=1$.
The wave-function becomes 
\bea\label{eq:wfss1}
\ket{\Psi(t)} = \sum_{I=1}^{N_G}\sum_{s=1}^{N_s} C_{I}^{(s)}(t)\ket{G_{I}(t)} \ket{\phi_{s}}
\eea
or equivalently
\bea\label{eq:wfss}
\ket{\Psi(t)} = \sum_{I=1}^{N_G}  \ket{G_{I}(t)} \left[\sum_{s=1}^{N_s} C_{I}^{(s)}(t) \ket{\phi_{s}}\right].
\eea
Therefore, this basis, also known as the single-set (SS) basis,  
can be either thought as consisting of stacks of identical gaussians replicated for all electronic states [\eq{eq:wfss1}] 
or gaussians with individual time-dependent electronic functions [\eq{eq:wfss}].
Although the SS gaussians always can exchange the population 
between electronic states, a new problem arises, replicas cannot take individual paths or decohere, instead 
each SS gaussian moves on an average, ``Ehrenfest-like" surface. To introduce more freedom, 
the cloning technique was suggested,\cite{Makhov:2014/jcp/054110} 
the algorithm monitors difference in forces that replicas within an SS stack 
experience on different electronic states. When the force difference becomes large, 
the cloning scheme splits the stack of gaussians
in two clones and adds empty replicas of gaussians for parts of the stack that went to another clone (\fig{fig:spcl}C). 
As in the case of spawning, cloning has been introduced for frozen-width gaussians that are evolving 
classically on  ``Ehrenfest-like" surfaces.\cite{Makhov:2014/jcp/054110,Fernandez:2016/pccp/10028} 
The algorithm treats every stack of gaussians 
independently, and thus, evaluation of forces is
straightforward. However, such cloning has the same drawback as spawning:
it does not treat each gaussian as a part of the total wave-function.

In order to
put the cloning idea on a rigorous quantum basis as well as to extend it to fully quantum treatment 
of nuclear dynamics one should consider the case of quantum entangled gaussians
with corresponding quantum forces that originate from the total nuclear wave-function. This is exactly 
the aim of the current Letter, where we propose a cloning algorithm for the fully quantum nonadiabatic
dynamics in the basis of SS frozen-width gaussians.

For the sake of simplicity, our technique will be illustrated on a set of two-state low dimensional 
diabatic models where the exact quantum results can be easily obtained. 
However, nothing prevents the use of the approach in the 
adiabatic representation with the on-the-fly generation of potential electronic surfaces.
To treat challenging geometric phase effects arising in the conical intersection case\cite{Mead:1979/jcp/2284,Ryabinkin:2013/prl/220406,Loic:2013/jcp/234103,Ryabinkin:2014/jcp/214116,Guo:2016/jacs}
 one can use recently introduced scheme evaluating adiabatic electronic functions only at gaussian centres.\cite{JoubertDoriol:2017im,Makhov:2014/jcp/054110,Fernandez:2016/pccp/10028}


\paragraph{Equations of motion for the SS representation:}
We start with the total non-stationary wave-function given by \eq{eq:wfss} where 
gaussians are taken in the coherent state (CS) form
\bea\notag
\langle \xx \ket{G_{I}(t)} &=& \prod_{j=1}^{N_{\rm dim}} \left(\frac{\omega_j}{\pi}\right)^{1/4} 
\exp\Big{[} -\frac{\omega_j}{2}[x_j -q_{jI}(t)]^2  \\ \label{eq:CS}
&&+ip_{jI}[x_j -q_{jI}(t)]+\frac{i}{2} p_{jI}q_{jI} \Big{]}
\eea
here, $\xx$ are nuclear coordinates, ${\rm dim}(\xx) = N_{\rm dim}$, and $\qq_I(t) = \{q_{jI}(t)\}_{j=1,N_{\rm dim}}$ and 
 $\pp_I(t)=\{p_{jI}(t)\}_{j=1,N_{\rm dim}}$ are time-dependent positions and 
 momenta.
EOM for all parameters of the wave-function in \eq{eq:wfss} can be obtained 
by finding an extremum of the action $S = \int \bra{\Psi(t)} \hat H - i\partial_t\ket{\Psi(t)} dt$ 
which is equivalent to solving\cite{Broeckhove:1988wo}
 \bea\label{eq:dL}
 {\rm Re}\bra{\delta\Psi(t)} \hat H - i\partial_t\ket{\Psi(t)} = 0.
 \eea
Here, $\hat H$ is the system Hamiltonian.
 For the parametrization of \eq{eq:wfss} it is easy to show that such version of TDVP 
 is equivalent to those of Dirac-Frenkel\cite{Dirac:1958/Book,Book/Frenkel:1934}
 and McLachlan\cite{McLachlan:1964ky}.
Introducing the SS variations
 \bea\label{eq:var1}
\ket{\delta\Psi} = \sum_{I,s} \Ket{\frac{\partial\Psi}{\partial C_I^{(s)}}}\delta C_I^{(s)} +
\sum_{j,I} \Ket{\frac{\partial\Psi}{\partial q_{jI}}} \delta q_{jI} + \Ket{\frac{\partial\Psi}{\partial p_{jI}}} \delta p_{jI}
\eea
and accounting for independence and arbitrariness of individual variations 
$\delta C_I^{(s)}$,  $\delta q_{jI}$, and $\delta p_{jI}$
leads to the EOM for all parameters\cite{Worth:2008/MP/2077,Richings:2015bn}
\bea\label{eq:cdot}
i\dot{C}_I^{(s)} &=& \sum_{J,s'}\Big{[}\mathbf{S^{-1}}(\mathbf{H_{ss'}}-i\boldsymbol{\tau}\delta_{ss'})\Big{]}_{IJ}C_{J}^{(s')}, \\
\label{eq:xidot}
i\dot{\xi}_{jI} &=& [\mathbf{B}^{-1} \mathbf{Y}]_{jI}, 
\eea
where $\xi_{jI}  = \omega_j q_{jI} +ip_{jI}$ are convenient variables encoding both position and momentum
components of CSs. Matrices involved in \eqs{eq:cdot} and \eqref{eq:xidot} are  
\bea
\tau_{IJ} &=& \bra{G_{I}}\partial_t G_{J}\rangle,~ S_{IJ} = \bra{G_{I}}G_{J}\rangle, \\
H_{ss',IJ} &=& \bra{G_I}\bra{\phi_s} \hat H \ket{\phi_{s'}}\ket{G_J}, \\
B_{Ik,Jn} &=& \sum_{s} C_I^{(s)*}C_J^{(s)} (\mathbf{S}^{(kn)} - 
\mathbf{S}^{(k0)}\mathbf{S}^{-1}\mathbf{S}^{(0n)})_{IJ}\\
Y_{Ik} &=& \sum_{ss',J} C_I^{(s)*}C_J^{(s')} (\mathbf{H}_{ss'}^{(k0)} - \mathbf{S}^{(k0)}\mathbf{S}^{-1}\mathbf{H}_{ss'})_{IJ}, \\	 
H_{ss',IJ}^{(k0)} &=& \Bra{\frac{\partial G_I}{\partial \xi_{kI}}} \bra{\phi_s} \hat H \ket{\phi_{s'}}\Big{\vert}G_J\Big{\rangle},~
S_{IJ}^{(kn)} = \Braket{\frac{\partial G_I}{\partial \xi_{kI}}\Big{\vert} \frac{\partial G_J}{\partial \xi_{nJ}}},~ \\
S_{IJ}^{(k0)} &=& \Braket{\frac{\partial G_I}{\partial \xi_{kI}}\Big{\vert} G_J},~ 
S_{IJ}^{(0n)} = \Braket{G_I\Big{\vert} \frac{\partial G_J}{\partial \xi_{nJ}}}.
\eea
Time-derivatives of CSs needed in the $\boldsymbol{\tau}$ matrix are
derived using the chain rule
\bea
\ket{\partial_t G_{K}} &=&  \Ket{\frac{\partial G_{K}}{\partial \qq_{K}}} 
\dot{\qq}_{K}(t) + \Ket{\frac{\partial G_{K}}{\partial \pp_{K}}}\dot{\pp}_{K}(t).
\eea
Solving equations \eqref{eq:cdot} and \eqref{eq:xidot} constitutes the vMCG approach within the SS basis set. 

\paragraph{Cloning SS pairs:}
If we consider a variation of the total wave-function that changes positions and momenta of replicas for an $I^{th}$ 
CS on different electronic states independently
 \bea\label{eq:var2}
\ket{\delta_I\Psi} = \sum_{s,j} \Ket{\frac{\partial\Psi}{\partial \xi_{jI}^{(s)}}} \delta \xi_{jI}^{(s)},
\eea
the condition of \eq{eq:dL} will not be satisfied. 
Formally, to consider such variation we need to evaluate it on a wave-function 
obtained from $\ket{\Psi}$ by allowing the $I^{th}$ CS's replicas to 
be different for different electronic states
\bea\label{eq:wfmix}
\ket{\Psi_I(t)} = \sum_s \left[C_{I}^{(s)}(t)\ket{G_{I}^{(s)}(t)}+ \sum_{J\ne I} C_{J}^{(s)}(t)  
\ket{G_{J}(t)} \right] \ket{\phi_{s}}.
\eea
To determine when and which of the SS pairs to split for cloning, it is instructive to consider the variation  
\bea
{\rm Re}\bra{\delta_I \Psi_I} \hat H-i\partial_t\ket{\Psi_I} &=&  {\rm Re}\left[\sum_{j,s} \delta\xi_{jI}^{(s)}
\Braket{\frac{\partial\Psi_I}{\partial \xi_{jI}^{(s)}}\Bigg{\vert} \hat H
-i\partial_t\Bigg{\vert}\Psi_I}\right] \neq 0.
\eea
This quantity contains arbitrary variations $\delta\xi_{jI}^{(s)}$, which can be removed 
if one is interested in effect of splitting of the $I^{th}$ SS pair on the action.
Thus, our criterion for splitting the $I^{th}$ SS pair is
\bea\label{eq:vd}
\sum_{j,s} \left\vert{\rm Re}\Braket{\frac{\partial\Psi_I}{\partial \xi_{jI}^{(s)}} 
\Bigg{\vert} \hat H-i\partial_t \Bigg{\vert}\Psi_I}\right\vert > \varepsilon
\eea
where $\varepsilon$ is an accuracy threshold. Interestingly, since we use 
$\ket{\Psi_I}$ from the SS simulation, the sum of derivatives over electronic states is always zero,
\bea
\sum_{s} {\rm Re}\Braket{\frac{\partial\Psi_I}{\partial \xi_{jI}^{(s)}} 
\Bigg{\vert} \hat H-i\partial_t \Bigg{\vert}\Psi_I}=0,
\eea
which is consistent with zero state average value.  
Therefore, the sum in \eq{eq:vd} corresponds to the norm of the deviation of generalized quantum 
state specific forces acting on an individual CS from the state averaged counterpart.

Note, that more than one SS pair can be split using the criterion of \eq{eq:vd} at a time, 
but for the sake of simplicity of further discussion we assume that only one pair has been split.   
Once the decision on splitting is made, to avoid linear dependency between clones, 
we propagate the split pair treated as independent CSs along with $N_G-1$ unsplit SS pairs. 
EOM for such hybrid evolution are obtained using TDVP applied 
to the parametrization $\Psi_I$ in \eq{eq:wfmix} and detailed in the SI.
CSs of the split pair move on different potential energy surfaces and necessarily
decohere so that the overlap integral $\bra{G_I^{(1)}} G_I^{(2)}\rangle$ will decrease allowing to create 
two new SS pairs without introducing linear dependency
\bea
  \begin{pmatrix} 
   C_I^{(1)}\ket{G_I^{(1)}} \\
  C_I^{(2)}\ket{G_I^{(2)}}
\end{pmatrix}
  \rightarrow
    \begin{pmatrix} 
    C_I^{(1)} \\
    0
  \end{pmatrix} \ket{G_I^{(1)}},~
      \begin{pmatrix} 
    0 \\
    C_I^{(2)}
 \end{pmatrix}\ket{G_I^{(2)}}
\eea
where vectors are written in the basis of electronic functions $\{\phi_s\}_{s=1,2}$. 
Once the split CSs are cloned into two new SS pairs the regular 
EOM (\eqref{eq:cdot} and \eqref{eq:xidot}) 
for $N_G+1$ SS pairs are employed. We will refer to this algorithm as the quantum cloning vMCG 
(QC-vMCG) approach.


\paragraph{Numerical examples:}
We illustrate the performance of QC-vMCG in modelling nuclear dynamics of one- and two-dimensional
two-state diabatic models 
\bea
  \label{eq:h}
 \hat H &=&   \sum_{j=1}^{N_{\rm dim}} 
  \begin{pmatrix} 
    [\hat p_j^2+\omega_j^2 {x}_j^2]/2 & c_j x_j \\
    c_j x_j & [\hat p_j^2+\omega_j^2 ({x}_j-d_j)^2]/2
  \end{pmatrix} +\begin{pmatrix} 
    0 & V \\
    V & \Delta
  \end{pmatrix},
\eea
where $x_j$ and $\hat p_j$ are nuclear coordinates and associated momenta, and 
$V,~\Delta,~d_j,~c_j,~\omega_j$ are constants. In the one-dimensional (1D) model ($N_{\rm dim}=1$), 
which is also known as spin-boson, $\hat H_{\rm SB}= \hat H$ where $c_j=0$. 
In the two-dimensional (2D) model ($N_{\rm dim}=2$), $\hat H_{\rm CI} = \hat H$ 
where $V=d_2=c_1=0$, this setup gives rise to the conical intersection of potential energies 
if transformed to the adiabatic representation. Other parameters in both cases are $\omega_1=0.89$,
$\omega_2=0.9$, $d_1=5$, and $\Delta=-\omega_1$. The last condition ensures 
resonance between vibrational levels of diabats coupled with linear potential coupling. 
Such resonances are unavoidable in large dimensional problems with conical intersections
but can be missing in 2D models.
We consider systems with strong and weak couplings, which are characterized by $V$ and $c_2$ 
for $\hat H_{\rm SB}$ and $\hat H_{\rm CI}$, respectively. Weak couplings simulate diabatically 
trapped systems,\cite{Izmaylov:2011ev} 
while strong couplings bring systems closer to the adiabatic limit. However, 
strong nonadiabatic couplings are present in both cases. Also, large reorganization energy is 
maintained in all systems ($d_1=5$) to make them challenging for the SS basis. 
Note that $d_j=c_j=0$  case can be solved with a single SS pair because diabatic states have 
identical nuclear dependence in this limit. 

\begin{figure}
  \centering
  \includegraphics[width=1\textwidth]{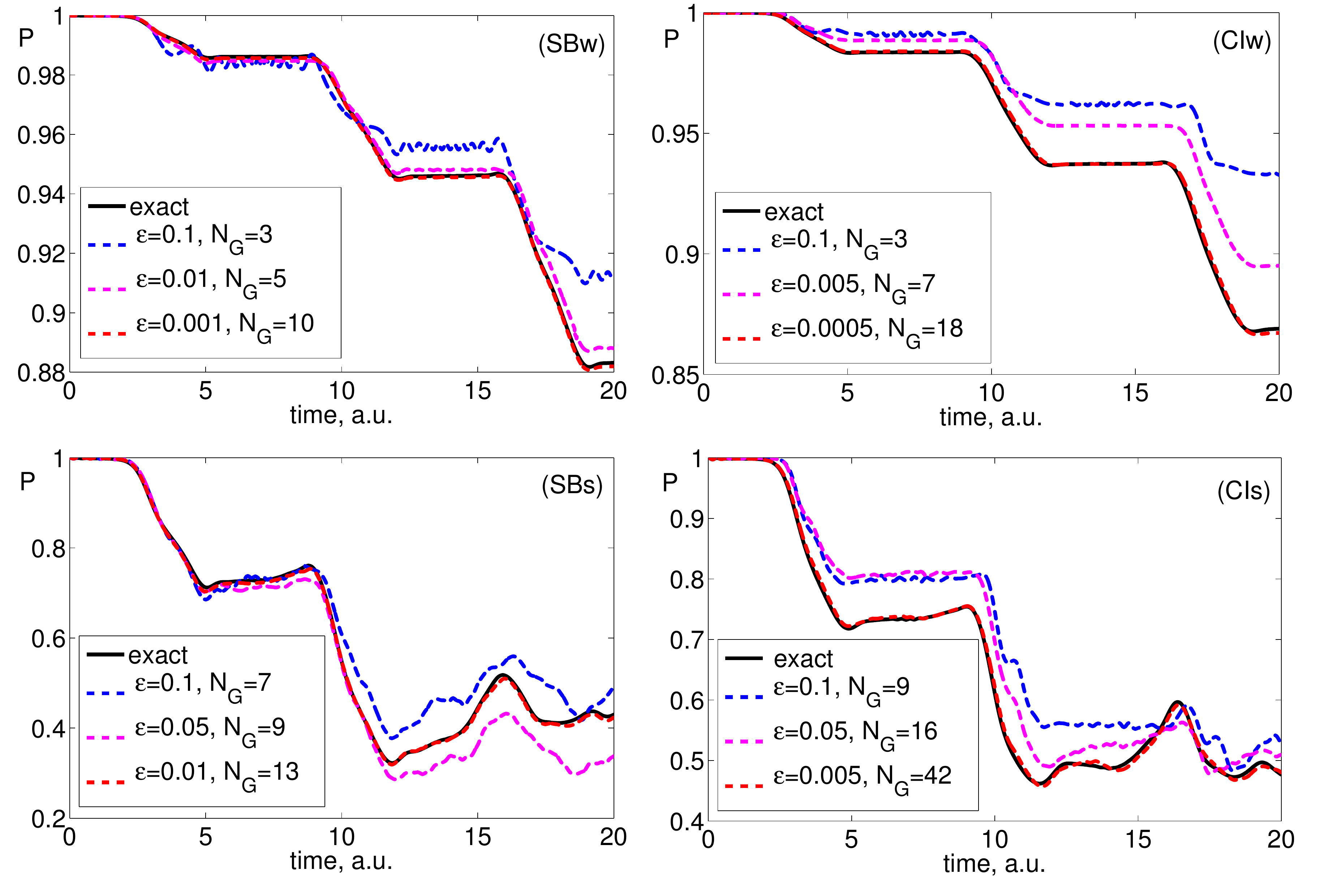}
  \caption{Electronic population as a function of time for $\hat H_{\rm SB}$ and $\hat H_{\rm CI}$  models for different couplings: (SBw)  $\hat H_{\rm SB}$ weak coupling, $V=0.1\omega_2$;  
  (SBs) $\hat H_{\rm SB}$ strong coupling, $V=0.5\omega_2$;  
  (CIw) $\hat H_{\rm CI}$ weak coupling, $c_2=0.1\omega_2$;  
  (CIs) $\hat H_{\rm CI}$ strong coupling, $c_2=0.5\omega_2$.
  In QC-vMCG (dashed curves) various $\varepsilon$'s produced different numbers of SS pairs at the end of the propagation, they are given by the $N_G$ values. 
 Positions of the initial CS are $q_c=-3$ (1D) and $\mathbf{q_c}=(-3,-1)$ (2D).}
  \label{fig:sbci}
\end{figure}

We simulate nuclear dynamics starting with an initial wave-function 
constituting a single SS pair 
$\ket{\Psi(t=0)} = \ket{G_1} [1\cdot\ket{\phi_1}+ 0\cdot\ket{\phi_2}]$ with zero initial 
momentum and centred at point $\mathbf{q_c}$. 
For each Hamiltonian we simulate time-dependent wave-functions and monitor 
the population of the 1st electronic state, $P(t) = {\rm Tr}_n[|\langle \phi_1\ket{\Psi(t)}|^2]$, 
where ${\rm Tr}_n$ is the trace over the nuclear coordinates (\fig{fig:sbci}). 
In all QC-vMCG calculations lowering $\varepsilon$ allowed us to converge to the exact dynamics 
generated by the split operator approach.\cite{Tannor:2007/214}
It may seem that lower couplings require 
lower thresholds, but it partly comes from the scale of the plots. Stronger couplings make 
initially empty replicas of CSs to be populated faster and to generate force difference
for faster decoherence.
Compare to 1D, in 2D there are more ways for CSs to avoid each other
and to lower the overlap between different pairs. Mutual help of CSs is weaker in 2D, and thus, 
more CSs are required in 2D for convergence. 

Besides cases in \fig{fig:sbci}, decoherence forces of \eq{eq:vd} for 
extreme limits of the spin-boson model have been considered: 
1) $\hat H_{\rm SB}$ with $d_1=0$ produces zero 
derivatives in \eq{eq:vd} because diabatic surfaces are parallel; 
2) $\hat H_{\rm SB}$ with $V\rightarrow 0$ 
also produces zero derivatives in \eq{eq:vd} because the population transfer is negligible, and the initial 
CSs evolves on a single harmonic oscillator.  
Also, in general case, it was confirmed that upon splitting not only derivatives in \eq{eq:vd} 
of the split pair vanish but also derivatives of unsplit CSs are reduced. The latter 
is the effect that comes from quantum entanglement of CSs in the total nuclear wave-function.  

In conclusion, we have introduced a novel general algorithm to extend basis set when needed 
in quantum dynamical simulations based on the magnitude of the 
quantum forces that have maximal effect on the action variation. 
These derivatives become large when pairs of nuclear CSs
located on different potential energy surfaces are experiencing very different forces. 
Similar developments were done for classically moving CSs, where an {\it ad hoc} criteria of pair separation 
were introduced based on force differences. We rigorously extended these intuitive techniques to 
fully quantum dynamics of CSs, with account for entanglement between different CSs 
in the total wave-function. 
Our approach can be easily extended to more than two electronic states, the adiabatic representation, 
and on-the-fly generation of potential energy surfaces. Another useful extension can be a 
formulation of a spawning technique which will use similar derivatives to determine 
a spawning event variationally. The work on this approach is underway and will be reported elsewhere.


{\it Acknowledgement:}
  A.F.I. thanks I. G. Ryabinkin for critical reading of the manuscript and 
  acknowledges funding from a Sloan Research Fellowship and the
  Natural Sciences and Engineering Research Council of Canada (NSERC)
  through the Discovery Grants Program.


\begin{thebibliography}{28}%
\makeatletter
\providecommand \@ifxundefined [1]{%
 \@ifx{#1\undefined}
}%
\providecommand \@ifnum [1]{%
 \ifnum #1\expandafter \@firstoftwo
 \else \expandafter \@secondoftwo
 \fi
}%
\providecommand \@ifx [1]{%
 \ifx #1\expandafter \@firstoftwo
 \else \expandafter \@secondoftwo
 \fi
}%
\providecommand \natexlab [1]{#1}%
\providecommand \enquote  [1]{``#1''}%
\providecommand \bibnamefont  [1]{#1}%
\providecommand \bibfnamefont [1]{#1}%
\providecommand \citenamefont [1]{#1}%
\providecommand \href@noop [0]{\@secondoftwo}%
\providecommand \href [0]{\begingroup \@sanitize@url \@href}%
\providecommand \@href[1]{\@@startlink{#1}\@@href}%
\providecommand \@@href[1]{\endgroup#1\@@endlink}%
\providecommand \@sanitize@url [0]{\catcode `\\12\catcode `\$12\catcode
  `\&12\catcode `\#12\catcode `\^12\catcode `\_12\catcode `\%12\relax}%
\providecommand \@@startlink[1]{}%
\providecommand \@@endlink[0]{}%
\providecommand \url  [0]{\begingroup\@sanitize@url \@url }%
\providecommand \@url [1]{\endgroup\@href {#1}{\urlprefix }}%
\providecommand \urlprefix  [0]{URL }%
\providecommand \Eprint [0]{\href }%
\providecommand \doibase [0]{http://dx.doi.org/}%
\providecommand \selectlanguage [0]{\@gobble}%
\providecommand \bibinfo  [0]{\@secondoftwo}%
\providecommand \bibfield  [0]{\@secondoftwo}%
\providecommand \translation [1]{[#1]}%
\providecommand \BibitemOpen [0]{}%
\providecommand \bibitemStop [0]{}%
\providecommand \bibitemNoStop [0]{.\EOS\space}%
\providecommand \EOS [0]{\spacefactor3000\relax}%
\providecommand \BibitemShut  [1]{\csname bibitem#1\endcsname}%
\let\auto@bib@innerbib\@empty
\bibitem [{\citenamefont {Kramer}\ and\ \citenamefont
  {Saraceno}(1981)}]{Book/Kramer:1981}%
  \BibitemOpen
  \bibfield  {author} {\bibinfo {author} {\bibfnamefont {P.}~\bibnamefont
  {Kramer}}\ and\ \bibinfo {author} {\bibfnamefont {M.}~\bibnamefont
  {Saraceno}},\ }\href@noop {} {\emph {\bibinfo {title} {Geometry of the
  Time-Dependent Variational Principle in Quantum Mechanics}}}\ (\bibinfo
  {publisher} {Springer},\ \bibinfo {address} {New York},\ \bibinfo {year}
  {1981})\BibitemShut {NoStop}%
\bibitem [{\citenamefont {Dirac}(1958)}]{Dirac:1958/Book}%
  \BibitemOpen
  \bibfield  {author} {\bibinfo {author} {\bibfnamefont {P.~A.~M.}\
  \bibnamefont {Dirac}},\ }\href@noop {} {\emph {\bibinfo {title} {The
  Principles of Quantum Mechanics, 4th Edition}}}\ (\bibinfo  {publisher}
  {Clarendon Press},\ \bibinfo {address} {Oxford},\ \bibinfo {year}
  {1958})\BibitemShut {NoStop}%
\bibitem [{\citenamefont {Frenkel}(1934)}]{Book/Frenkel:1934}%
  \BibitemOpen
  \bibfield  {author} {\bibinfo {author} {\bibfnamefont {J.}~\bibnamefont
  {Frenkel}},\ }\href@noop {} {\emph {\bibinfo {title} {Wave Mechanics}}}\
  (\bibinfo  {publisher} {Clarendon Press},\ \bibinfo {address} {Oxford},\
  \bibinfo {year} {1934})\BibitemShut {NoStop}%
\bibitem [{\citenamefont {Meyer}, \citenamefont {Manthe},\ and\ \citenamefont
  {Cederbaum}(1990)}]{mey90:73}%
  \BibitemOpen
  \bibfield  {author} {\bibinfo {author} {\bibfnamefont {H.-D.}\ \bibnamefont
  {Meyer}}, \bibinfo {author} {\bibfnamefont {U.}~\bibnamefont {Manthe}}, \
  and\ \bibinfo {author} {\bibfnamefont {L.~S.}\ \bibnamefont {Cederbaum}},\
  }\href@noop {} {\bibfield  {journal} {\bibinfo  {journal} {Chem.\ Phys.\
  Lett.}\ }\textbf {\bibinfo {volume} {165}},\ \bibinfo {pages} {73} (\bibinfo
  {year} {1990})}\BibitemShut {NoStop}%
\bibitem [{\citenamefont {Wang}\ and\ \citenamefont
  {Thoss}(2003{\natexlab{a}})}]{Wang:2003/jcp/1289}%
  \BibitemOpen
  \bibfield  {author} {\bibinfo {author} {\bibfnamefont {H.}~\bibnamefont
  {Wang}}\ and\ \bibinfo {author} {\bibfnamefont {M.}~\bibnamefont {Thoss}},\
  }\href {\doibase 10.1063/1.1580111} {\bibfield  {journal} {\bibinfo
  {journal} {J. Chem. Phys.}\ }\textbf {\bibinfo {volume} {119}},\ \bibinfo
  {pages} {1289} (\bibinfo {year} {2003}{\natexlab{a}})}\BibitemShut {NoStop}%
\bibitem [{\citenamefont {G.~A.~Worth}\ and\ \citenamefont {Meyer}()}]{mctdh}%
  \BibitemOpen
  \bibfield  {author} {\bibinfo {author} {\bibfnamefont {A.~J.}\ \bibnamefont
  {G.~A.~Worth}, \bibfnamefont {M.~H.~Beck}}\ and\ \bibinfo {author}
  {\bibfnamefont {H.-D.}\ \bibnamefont {Meyer}},\ }\href@noop {} {}\bibinfo
  {note} {{\em The MCTDH Package,} Development Version 9.0, University of
  Heidelberg, Heidelberg, Germany, 2009.}\BibitemShut {Stop}%
\bibitem [{\citenamefont {Wang}\ and\ \citenamefont
  {Thoss}(2003{\natexlab{b}})}]{Wang:2003fu}%
  \BibitemOpen
  \bibfield  {author} {\bibinfo {author} {\bibfnamefont {H.}~\bibnamefont
  {Wang}}\ and\ \bibinfo {author} {\bibfnamefont {M.}~\bibnamefont {Thoss}},\
  }\href@noop {} {\bibfield  {journal} {\bibinfo  {journal} {The Journal of
  Chemical Physics}\ }\textbf {\bibinfo {volume} {119}},\ \bibinfo {pages}
  {1289} (\bibinfo {year} {2003}{\natexlab{b}})}\BibitemShut {NoStop}%
\bibitem [{\citenamefont {Manthe}(2008)}]{Manthe:2008ev}%
  \BibitemOpen
  \bibfield  {author} {\bibinfo {author} {\bibfnamefont {U.}~\bibnamefont
  {Manthe}},\ }\href@noop {} {\bibfield  {journal} {\bibinfo  {journal} {The
  Journal of Chemical Physics}\ }\textbf {\bibinfo {volume} {128}},\ \bibinfo
  {pages} {164116} (\bibinfo {year} {2008})}\BibitemShut {NoStop}%
\bibitem [{\citenamefont {Yang}\ \emph {et~al.}(2009)\citenamefont {Yang},
  \citenamefont {Coe}, \citenamefont {Kaduk},\ and\ \citenamefont
  {Mart{\'\i}nez}}]{Yang:2009ja}%
  \BibitemOpen
  \bibfield  {author} {\bibinfo {author} {\bibfnamefont {S.}~\bibnamefont
  {Yang}}, \bibinfo {author} {\bibfnamefont {J.~D.}\ \bibnamefont {Coe}},
  \bibinfo {author} {\bibfnamefont {B.}~\bibnamefont {Kaduk}}, \ and\ \bibinfo
  {author} {\bibfnamefont {T.~J.}\ \bibnamefont {Mart{\'\i}nez}},\ }\href@noop
  {} {\bibfield  {journal} {\bibinfo  {journal} {J. Chem. Phys.}\ }\textbf
  {\bibinfo {volume} {130}},\ \bibinfo {pages} {134113} (\bibinfo {year}
  {2009})}\BibitemShut {NoStop}%
\bibitem [{\citenamefont {Ben-Nun}\ and\ \citenamefont
  {Martinez}(2002)}]{BenNun:2002tx}%
  \BibitemOpen
  \bibfield  {author} {\bibinfo {author} {\bibfnamefont {M.}~\bibnamefont
  {Ben-Nun}}\ and\ \bibinfo {author} {\bibfnamefont {T.~J.}\ \bibnamefont
  {Martinez}},\ }\href@noop {} {\bibfield  {journal} {\bibinfo  {journal} {Adv.
  Chem. Phys.}\ }\textbf {\bibinfo {volume} {121}},\ \bibinfo {pages} {439}
  (\bibinfo {year} {2002})}\BibitemShut {NoStop}%
\bibitem [{\citenamefont {Shalashilin}(2009)}]{Shalashilin:2009/JCP/244101}%
  \BibitemOpen
  \bibfield  {author} {\bibinfo {author} {\bibfnamefont {D.~V.}\ \bibnamefont
  {Shalashilin}},\ }\href@noop {} {\bibfield  {journal} {\bibinfo  {journal}
  {J. Chem. Phys.}\ }\textbf {\bibinfo {volume} {130}},\ \bibinfo {pages}
  {244101} (\bibinfo {year} {2009})}\BibitemShut {NoStop}%
\bibitem [{\citenamefont {Burghardt}, \citenamefont {Giri},\ and\ \citenamefont
  {Worth}(2008)}]{Burghardt:2008iz}%
  \BibitemOpen
  \bibfield  {author} {\bibinfo {author} {\bibfnamefont {I.}~\bibnamefont
  {Burghardt}}, \bibinfo {author} {\bibfnamefont {K.}~\bibnamefont {Giri}}, \
  and\ \bibinfo {author} {\bibfnamefont {G.~A.}\ \bibnamefont {Worth}},\
  }\href@noop {} {\bibfield  {journal} {\bibinfo  {journal} {J. Chem. Phys.}\
  }\textbf {\bibinfo {volume} {129}},\ \bibinfo {pages} {174104} (\bibinfo
  {year} {2008})}\BibitemShut {NoStop}%
\bibitem [{\citenamefont {Worth}, \citenamefont {Robb},\ and\ \citenamefont
  {Lasorne}(2008)}]{Worth:2008/MP/2077}%
  \BibitemOpen
  \bibfield  {author} {\bibinfo {author} {\bibfnamefont {G.~A.}\ \bibnamefont
  {Worth}}, \bibinfo {author} {\bibfnamefont {M.~A.}\ \bibnamefont {Robb}}, \
  and\ \bibinfo {author} {\bibfnamefont {B.}~\bibnamefont {Lasorne}},\
  }\href@noop {} {\bibfield  {journal} {\bibinfo  {journal} {Mol. Phys.}\
  }\textbf {\bibinfo {volume} {106}},\ \bibinfo {pages} {2077} (\bibinfo {year}
  {2008})}\BibitemShut {NoStop}%
\bibitem [{\citenamefont {Worth}, \citenamefont {Robb},\ and\ \citenamefont
  {Burghardt}(2004)}]{Worth:2004/FD/307}%
  \BibitemOpen
  \bibfield  {author} {\bibinfo {author} {\bibfnamefont {G.~A.}\ \bibnamefont
  {Worth}}, \bibinfo {author} {\bibfnamefont {M.~A.}\ \bibnamefont {Robb}}, \
  and\ \bibinfo {author} {\bibfnamefont {I.}~\bibnamefont {Burghardt}},\
  }\href@noop {} {\bibfield  {journal} {\bibinfo  {journal} {Faraday Discuss.}\
  }\textbf {\bibinfo {volume} {127}},\ \bibinfo {pages} {307} (\bibinfo {year}
  {2004})}\BibitemShut {NoStop}%
\bibitem [{\citenamefont {Izmaylov}(2013)}]{Izmaylov:2013fe}%
  \BibitemOpen
  \bibfield  {author} {\bibinfo {author} {\bibfnamefont {A.~F.}\ \bibnamefont
  {Izmaylov}},\ }\href@noop {} {\bibfield  {journal} {\bibinfo  {journal} {J.
  Chem. Phys.}\ }\textbf {\bibinfo {volume} {138}},\ \bibinfo {pages} {104115}
  (\bibinfo {year} {2013})}\BibitemShut {NoStop}%
\bibitem [{\citenamefont {Makhov}\ \emph {et~al.}(2014)\citenamefont {Makhov},
  \citenamefont {Glover}, \citenamefont {Martinez},\ and\ \citenamefont
  {Shalashilin}}]{Makhov:2014/jcp/054110}%
  \BibitemOpen
  \bibfield  {author} {\bibinfo {author} {\bibfnamefont {D.~V.}\ \bibnamefont
  {Makhov}}, \bibinfo {author} {\bibfnamefont {W.~J.}\ \bibnamefont {Glover}},
  \bibinfo {author} {\bibfnamefont {T.~J.}\ \bibnamefont {Martinez}}, \ and\
  \bibinfo {author} {\bibfnamefont {D.~V.}\ \bibnamefont {Shalashilin}},\
  }\href {\doibase 10.1063/1.4891530} {\bibfield  {journal} {\bibinfo
  {journal} {J. Chem. Phys.}\ }\textbf {\bibinfo {volume} {141}},\ \bibinfo
  {pages} {054110} (\bibinfo {year} {2014})}\BibitemShut {NoStop}%
\bibitem [{\citenamefont {Fernandez-Alberti}\ \emph {et~al.}(2016)\citenamefont
  {Fernandez-Alberti}, \citenamefont {Makhov}, \citenamefont {Tretiak},\ and\
  \citenamefont {Shalashilin}}]{Fernandez:2016/pccp/10028}%
  \BibitemOpen
  \bibfield  {author} {\bibinfo {author} {\bibfnamefont {S.}~\bibnamefont
  {Fernandez-Alberti}}, \bibinfo {author} {\bibfnamefont {D.~V.}\ \bibnamefont
  {Makhov}}, \bibinfo {author} {\bibfnamefont {S.}~\bibnamefont {Tretiak}}, \
  and\ \bibinfo {author} {\bibfnamefont {D.~V.}\ \bibnamefont {Shalashilin}},\
  }\href {\doibase 10.1039/C5CP07332D} {\bibfield  {journal} {\bibinfo
  {journal} {Phys. Chem. Chem. Phys.}\ }\textbf {\bibinfo {volume} {18}},\
  \bibinfo {pages} {10028} (\bibinfo {year} {2016})}\BibitemShut {NoStop}%
\bibitem [{\citenamefont {Mead}\ and\ \citenamefont
  {Truhlar}(1979)}]{Mead:1979/jcp/2284}%
  \BibitemOpen
  \bibfield  {author} {\bibinfo {author} {\bibfnamefont {C.~A.}\ \bibnamefont
  {Mead}}\ and\ \bibinfo {author} {\bibfnamefont {D.~G.}\ \bibnamefont
  {Truhlar}},\ }\href {\doibase 10.1063/1.437734} {\bibfield  {journal}
  {\bibinfo  {journal} {J. Chem. Phys.}\ }\textbf {\bibinfo {volume} {70}},\
  \bibinfo {pages} {2284} (\bibinfo {year} {1979})}\BibitemShut {NoStop}%
\bibitem [{\citenamefont {Ryabinkin}\ and\ \citenamefont
  {Izmaylov}(2013)}]{Ryabinkin:2013/prl/220406}%
  \BibitemOpen
  \bibfield  {author} {\bibinfo {author} {\bibfnamefont {I.~G.}\ \bibnamefont
  {Ryabinkin}}\ and\ \bibinfo {author} {\bibfnamefont {A.~F.}\ \bibnamefont
  {Izmaylov}},\ }\href {\doibase 10.1103/PhysRevLett.111.220406} {\bibfield
  {journal} {\bibinfo  {journal} {Phys. Rev. Lett.}\ }\textbf {\bibinfo
  {volume} {111}},\ \bibinfo {pages} {220406} (\bibinfo {year}
  {2013})}\BibitemShut {NoStop}%
\bibitem [{\citenamefont {Joubert-Doriol}, \citenamefont {Ryabinkin},\ and\
  \citenamefont {Izmaylov}(2013)}]{Loic:2013/jcp/234103}%
  \BibitemOpen
  \bibfield  {author} {\bibinfo {author} {\bibfnamefont {L.}~\bibnamefont
  {Joubert-Doriol}}, \bibinfo {author} {\bibfnamefont {I.~G.}\ \bibnamefont
  {Ryabinkin}}, \ and\ \bibinfo {author} {\bibfnamefont {A.~F.}\ \bibnamefont
  {Izmaylov}},\ }\href {\doibase http://dx.doi.org/10.1063/1.4844095}
  {\bibfield  {journal} {\bibinfo  {journal} {J. Chem. Phys.}\ }\textbf
  {\bibinfo {volume} {139}},\ \bibinfo {pages} {234103} (\bibinfo {year}
  {2013})}\BibitemShut {NoStop}%
\bibitem [{\citenamefont {Ryabinkin}, \citenamefont {Joubert-Doriol},\ and\
  \citenamefont {Izmaylov}(2014)}]{Ryabinkin:2014/jcp/214116}%
  \BibitemOpen
  \bibfield  {author} {\bibinfo {author} {\bibfnamefont {I.~G.}\ \bibnamefont
  {Ryabinkin}}, \bibinfo {author} {\bibfnamefont {L.}~\bibnamefont
  {Joubert-Doriol}}, \ and\ \bibinfo {author} {\bibfnamefont {A.~F.}\
  \bibnamefont {Izmaylov}},\ }\href@noop {} {\bibfield  {journal} {\bibinfo
  {journal} {J. Chem. Phys.}\ }\textbf {\bibinfo {volume} {140}},\ \bibinfo
  {pages} {214116} (\bibinfo {year} {2014})}\BibitemShut {NoStop}%
\bibitem [{\citenamefont {Xie}\ \emph {et~al.}(2016)\citenamefont {Xie},
  \citenamefont {Ma}, \citenamefont {Zhu}, \citenamefont {Yarkony},
  \citenamefont {Xie},\ and\ \citenamefont {Guo}}]{Guo:2016/jacs}%
  \BibitemOpen
  \bibfield  {author} {\bibinfo {author} {\bibfnamefont {C.}~\bibnamefont
  {Xie}}, \bibinfo {author} {\bibfnamefont {J.}~\bibnamefont {Ma}}, \bibinfo
  {author} {\bibfnamefont {X.}~\bibnamefont {Zhu}}, \bibinfo {author}
  {\bibfnamefont {D.~R.}\ \bibnamefont {Yarkony}}, \bibinfo {author}
  {\bibfnamefont {D.}~\bibnamefont {Xie}}, \ and\ \bibinfo {author}
  {\bibfnamefont {H.}~\bibnamefont {Guo}},\ }\href@noop {} {\bibfield
  {journal} {\bibinfo  {journal} {J. Am. Chem. Soc.}\ }\textbf {\bibinfo
  {volume} {138}},\ \bibinfo {pages} {7828} (\bibinfo {year}
  {2016})}\BibitemShut {NoStop}%
\bibitem [{\citenamefont {Joubert-Doriol}\ \emph {et~al.}(2017)\citenamefont
  {Joubert-Doriol}, \citenamefont {Sivasubramanium}, \citenamefont
  {Ryabinkin},\ and\ \citenamefont {Izmaylov}}]{JoubertDoriol:2017im}%
  \BibitemOpen
  \bibfield  {author} {\bibinfo {author} {\bibfnamefont {L.}~\bibnamefont
  {Joubert-Doriol}}, \bibinfo {author} {\bibfnamefont {J.}~\bibnamefont
  {Sivasubramanium}}, \bibinfo {author} {\bibfnamefont {I.~G.}\ \bibnamefont
  {Ryabinkin}}, \ and\ \bibinfo {author} {\bibfnamefont {A.~F.}\ \bibnamefont
  {Izmaylov}},\ }\href@noop {} {\bibfield  {journal} {\bibinfo  {journal} {The
  Journal of Physical Chemistry Letters}\ ,\ \bibinfo {pages} {452}} (\bibinfo
  {year} {2017})}\BibitemShut {NoStop}%
\bibitem [{\citenamefont {Broeckhove}\ \emph {et~al.}(1988)\citenamefont
  {Broeckhove}, \citenamefont {Lathouwers}, \citenamefont {Kesteloot},\ and\
  \citenamefont {Van~Leuven}}]{Broeckhove:1988wo}%
  \BibitemOpen
  \bibfield  {author} {\bibinfo {author} {\bibfnamefont {J.}~\bibnamefont
  {Broeckhove}}, \bibinfo {author} {\bibfnamefont {L.}~\bibnamefont
  {Lathouwers}}, \bibinfo {author} {\bibfnamefont {E.}~\bibnamefont
  {Kesteloot}}, \ and\ \bibinfo {author} {\bibfnamefont {P.}~\bibnamefont
  {Van~Leuven}},\ }\href@noop {} {\bibfield  {journal} {\bibinfo  {journal}
  {Chemical Physics Letters}\ }\textbf {\bibinfo {volume} {149}},\ \bibinfo
  {pages} {547} (\bibinfo {year} {1988})}\BibitemShut {NoStop}%
\bibitem [{\citenamefont {McLachlan}(1964)}]{McLachlan:1964ky}%
  \BibitemOpen
  \bibfield  {author} {\bibinfo {author} {\bibfnamefont {A.~D.}\ \bibnamefont
  {McLachlan}},\ }\href@noop {} {\bibfield  {journal} {\bibinfo  {journal}
  {Molecular Physics}\ }\textbf {\bibinfo {volume} {8}},\ \bibinfo {pages} {39}
  (\bibinfo {year} {1964})}\BibitemShut {NoStop}%
\bibitem [{\citenamefont {Richings}\ \emph {et~al.}(2015)\citenamefont
  {Richings}, \citenamefont {Polyak}, \citenamefont {Spinlove}, \citenamefont
  {Worth}, \citenamefont {Burghardt},\ and\ \citenamefont
  {Lasorne}}]{Richings:2015bn}%
  \BibitemOpen
  \bibfield  {author} {\bibinfo {author} {\bibfnamefont {G.~W.}\ \bibnamefont
  {Richings}}, \bibinfo {author} {\bibfnamefont {I.}~\bibnamefont {Polyak}},
  \bibinfo {author} {\bibfnamefont {K.~E.}\ \bibnamefont {Spinlove}}, \bibinfo
  {author} {\bibfnamefont {G.~A.}\ \bibnamefont {Worth}}, \bibinfo {author}
  {\bibfnamefont {I.}~\bibnamefont {Burghardt}}, \ and\ \bibinfo {author}
  {\bibfnamefont {B.}~\bibnamefont {Lasorne}},\ }\href@noop {} {\bibfield
  {journal} {\bibinfo  {journal} {International Reviews in Physical Chemistry}\
  ,\ \bibinfo {pages} {161}} (\bibinfo {year} {2015})}\BibitemShut {NoStop}%
\bibitem [{\citenamefont {Izmaylov}\ \emph {et~al.}(2011)\citenamefont
  {Izmaylov}, \citenamefont {Mendive~Tapia}, \citenamefont {Bearpark},
  \citenamefont {Robb}, \citenamefont {Tully},\ and\ \citenamefont
  {Frisch}}]{Izmaylov:2011ev}%
  \BibitemOpen
  \bibfield  {author} {\bibinfo {author} {\bibfnamefont {A.~F.}\ \bibnamefont
  {Izmaylov}}, \bibinfo {author} {\bibfnamefont {D.}~\bibnamefont
  {Mendive~Tapia}}, \bibinfo {author} {\bibfnamefont {M.~J.}\ \bibnamefont
  {Bearpark}}, \bibinfo {author} {\bibfnamefont {M.~A.}\ \bibnamefont {Robb}},
  \bibinfo {author} {\bibfnamefont {J.~C.}\ \bibnamefont {Tully}}, \ and\
  \bibinfo {author} {\bibfnamefont {M.~J.}\ \bibnamefont {Frisch}},\
  }\href@noop {} {\bibfield  {journal} {\bibinfo  {journal} {The Journal of
  Chemical Physics}\ }\textbf {\bibinfo {volume} {135}},\ \bibinfo {pages}
  {234106} (\bibinfo {year} {2011})}\BibitemShut {NoStop}%
\bibitem [{\citenamefont {Tannor}(2007)}]{Tannor:2007/214}%
  \BibitemOpen
  \bibfield  {author} {\bibinfo {author} {\bibfnamefont {D.~J.}\ \bibnamefont
  {Tannor}},\ }in\ \href@noop {} {\emph {\bibinfo {booktitle} {Introduction to
  Quantum Mechanics: A Time-Dependent Perspective}}}\ (\bibinfo  {publisher}
  {University Science Books},\ \bibinfo {address} {Sausalito, California},\
  \bibinfo {year} {2007})\ p.\ \bibinfo {pages} {214}\BibitemShut {NoStop}%
\end{thebibliography}
%

\end{document}